\documentclass[%
 reprint,
amsmath,amssymb,
aps,
]{revtex4-1}

\usepackage{graphicx}
\usepackage{dcolumn}
\usepackage{bm}

\begin{document}
\preprint{APS/123-QED}

\title{Photoproduction of Upsilon 
in Ultraperipheral Collisions at LHC Run2 energies}

\author{Dipanwita Dutta}
\email{ddutta@barc.gov.in}
\affiliation{
Nuclear Physics Division, Bhabha Atomic Research Centre, Mumbai-400085, India
}%
\author{Ruchi Chudasama}%
\affiliation{
Nuclear Physics Division, Bhabha Atomic Research Centre, Mumbai-400085, India
}%

\date{\today}

\begin{abstract}
The exclusive $\Upsilon$ photoproduction in ultraperipheral 
proton-nucleus and nucleus-nucleus 
collisions at LHC energies is being investigated using pQCD
framework which constrain the gluon distribution in the proton and nuclei at low $x$. 
Rapidity distributions of 
$\Upsilon$ photoproduction with different gluon distribution parameterization
and gluon shadowing for nuclear PDFs including
photon flux suppression for strong interaction, are
being presented at LHC energies. Predictions done
for ultraperipheral collisions at 
$\sqrt s$=$8.16$ TeV in pPb,
and at  $\sqrt s$=$5.02$ TeV 
in PbPb collisions  which are LHC Run 2 Heavy Ion collision scenario
 and measurements of cross-section  
will be  available soon.    
\end{abstract}

\maketitle

\section{\label{sec:s1} Introduction}
Ultraperipheral collisions (UPCs) of protons (ions) 
corresponds to the scenario when the impact parameter is larger than
the sum of the radii of colliding hadron (ion), so that strong interaction
is suppressed and they interact electromagnetically via emission of
quasi-real photons. The large energies of the virtual
photons produced in such reactions has stimulated many studies
of photoproduction of heavy vector mesons
($J/\psi$ and $\Upsilon$(nS)) ~\cite{Baltz:2008bb,Enterria:2008hi,Brodsky:1999,Starlight:1999,Starlight:2004,
Frankfurt:2001,Strikman:2006sv,Motyka:2008,Lappi:2010,Lappi:2013,guzey01,Guzey:2013aa,Guzey:2013bb,Ryskin:2013jmr,Ryskin:2008rt,
khoze:2000,Khoze:2010,Adeluyi:2012aa,Adeluyi:2012,Adeluyi:2013,sampaio,Goncalves:2014,Goncalves:2017aa,Goncalves:2017,Silveira:2015,Silveira:2017},
where large mass of the quarks provide a hard scale,
justifying the use of factorization theorem of perturbative QCD (pQCD).
Within leading logarithmic approximation of pQCD,
the cross section of photoproduction of $J/\psi$ or $\Upsilon$
is proportional to the square of gluon density of the target.
As the fraction $x$  of target momentum carried by 
the gluon is inversely proportional to the
beam momentum, at LHC energies it is possible to explore the
small $x$ behavior of the gluon density in the proton
and nuclei. In particular, it provides a sensitive tool 
to probe the onset of gluon  saturation or nuclear 
gluon shadowing effects in small $x$.

Recently  LHCb collaboration measured the yield of $J/\psi$ 
and $\Upsilon$~\cite{lhcb01,LHCb:2015} at forward rapidities 
($2<y<4.5$) in proton-proton 
UPC at $7$, $8$ TeV corresponds to $x$ range for $J/\psi$, 
$6\times 10^{-6} < x < 6 \times 10^{-5}$ and 
down to $x \sim 1.5 \times 10^{-5}$ for $\Upsilon$(nS).
Though $J/\psi$ analysis confirmed the power law
energy dependence of the $\gamma p \rightarrow J/\psi p$
cross section consistent with HERA results
~\cite{H1:2000,H1:2006,H1:2013}, but
exclusive $\Upsilon$ measurement shows  preference
to the estimations including next-to-leading order calculation
in contrast to HERA results of $\Upsilon$ photoproduction
~\cite{herazeus,ZEUS:2009,ZEUS:2012}.

The ALICE collaboration measured the coherent $J/\psi$ photoproduction~\cite{alice01,alice02}
in ultraperipheral collisions of PbPb at $\sqrt s_{NN}=2.76$ TeV.
The cross section was measured in two regions of the rapidity
of produced $J/\psi$ : $-3.6 < y <-2.6$ and at central rapidities
$-0.9 < y <0.9$ which corresponds to 
$x=2\times 10^{-2}$ and $x\sim 10^{-3}$ respectively.
The results were compared with different model predictions
and shows good agreement with models including nuclear gluon shadowing.
ALICE also measured $J/\psi$ photoproduction~\cite{alice03}
in  pPb UPC at $\sqrt {s_{NN}}=5.02$ TeV
in the rapidity range $2.5 < y <4$ (p-Pb) or $-3.6 < y <-2.6$ (Pb-p)
corresponds to $x\sim 2\times 10^{-2}$ and $\sim 2\times 10^{-5}$ which
didn't show significant change in the gluon density behavior
of the proton between HERA and LHC energies.

CMS collaboration ~\cite{CMS:pPb,CMS:pPba} also measured 
 $\Upsilon(1S)$
photoproduction cross section with pPb collisions at $\sqrt {s_{NN}}=5.02$ TeV
in the central rapidity range $-2.2 < y <2.2$ 
corresponds to the Bjorken $x$ value $ 1.3 \times 10^{-4} < x <  10^{-2}$.
This result will provides the behavior of
$\Upsilon(1S)$ photoproduction cross section with
$W_{\gamma p}$, photon-proton center-of-mass energy,  in the region 
between HERA~\cite{herazeus,ZEUS:2009,ZEUS:2012} and
LHCb~\cite{LHCb:2015} experiment.

The aim of the paper is the following.
 We estimate
the $\Upsilon$ photoproduction in $pPb$ collisions
which is much preferable in comparison to proton-proton collision,
due to asymmetric nature. 
Using the leading order
(LO) pQCD, first we estimate  of $\gamma p \rightarrow \Upsilon(nS) p$
and $\gamma Pb \rightarrow \Upsilon(nS) Pb$
employing different parameterization of the gluon distribution
in the proton and nucleus and compare with the data of HERA and LHCb. 
Secondly, we calculate the rapidity distribution of exclusive
$\Upsilon$ photoproduction in the proton-lead
collisions  for $\sqrt s_{NN}=5.02$ TeV 
in the kinematic range of CMS experiment~\cite{Chatrchyan:2008zzk} and 
also done predictions  in pPb UPC for $\sqrt s_{NN}=8.16$ TeV
 which corresponds to the collision scenario of Run 2 LHC:
 data taken during Nov.-Dec. 2016. 
The suppression of photon flux
due to strong interaction, which can reach up to $\sim 20\%$ difference in photon
flux in forward rapidities~\cite{guzey01} in  proton-Pb UPC,
is included by modifying the photon flux using
the Glauber model of multiple proton-nucleus scattering.
We also made predictions for the $\Upsilon$ photoproduction in
PbPb UPC at $\sqrt s_{NN}=5.02$ TeV, the scenario of LHC Run2 : data
taken during Nov.-Dec. 2015,
taking into account of recently available nuclear gluon shadowing
parameterizations.

The paper is organized as follows. In section II, we discuss
the photon flux generated by the proton and the nucleus.
In section III, we discuss the photoproduction of $\Upsilon(1S)$
from proton with different gluon PDF  and 
comparison with data. In section IV, we produce the 
results of photoproduction cross section of $\Upsilon$  with rapidity and discuss
predictions for LHC Run 2.

\section{\label{sec:s2}Photon flux}
The key ingredient of this study
is the estimation of photon flux from proton and the lead
nucleus. In case of proton-nucleus UPCs one needs to take into account
the suppression of the flux due to the strong interaction between colliding particles.
The photon flux of the proton (nucleus) of charge Z
can be expressed as the convolution over the impact parameter
$b$~\cite{guzey01}:
\begin{eqnarray}
N_{\gamma/Z}(\omega)=\int_{0}^{\infty} d^{2}\vec {b}  ~\Gamma_{pA}
(\vec{b}) N_{\gamma/Z}(\omega,\vec{b})~, 
\label{eq:fluxexact}
\end{eqnarray}
 where $ N_{\gamma/Z}(\omega,\vec{b})$ is the photon flux 
in the transverse distance $\vec{ b}$ away from the proton (nucleus)~\cite{vidovic},  
\begin{eqnarray}
N_{\gamma/Z}(\omega,\vec{b})=C\left(\int_{0}^{\infty}dk_{\perp}
\frac{k_{\perp}^2F_Z(k_{\perp}^2+\omega^2/\gamma_L^2)}{k_{\perp}^2+\omega^2/\gamma_L^2}  
J_1(bk_{\perp})\right)^2,\nonumber\\
\label{eq:fluxim}
\end{eqnarray}
and $\Gamma_{pA}(\vec{b})$ is the probability to suppress
the proton-nucleus strong interaction at small impact parameter $b$~\cite{guzey01},
\begin{eqnarray}
~\Gamma_{pA}(\vec{b})=\exp \left( -\sigma_{NN}\int_{-\infty}^{\infty} dz \rho_A(z,\vec{b})\right). 
\label{eq:fluxsup}
\end{eqnarray}
In Eq.~\ref{eq:fluxim}, $C=Z^2\alpha_{em}/\pi^2$ where   
$\alpha_{em}$ is the fine structure constant, $F_{Z}(Q^2)$ is the 
charge form factor of the proton (nucleus), $\gamma_L$ is the Lorentz factor 
($\gamma_L=2670,~4340$ for pPb at $\sqrt s_{NN}= 5.02,~8.16$ TeV, 
 $\gamma_L=3730,~6920,~7450$ for
pp at $\sqrt s= 7,~13,~14$ TeV respectively and $\gamma_L=2670$ for PbPb at $\sqrt s_{NN}= 5.02$ TeV), 
$\omega$ is the energy of the emitted photon and $J_1$ is the Bessel function of the first kind.
In Eq.~\ref{eq:fluxsup}, $\sigma_{NN}$ is the total nucleon-nucleon cross section
and $\rho_A(\vec{r})$ is the nuclear density.

For $N_{\gamma/Z}(\omega)$ for the photon flux from the proton 
one generally use an approximate expression from Drees and Zeppenfeld~\cite{Drees89},
\begin{eqnarray}
N_{\gamma/p}(\omega)=&\frac{\alpha_{em}}{2\pi}\left[1+\left(1-\frac{2\omega}{\sqrt{s_{NN}}}\right)\right]\nonumber\\
&\left[\ln D-\frac{11}{6}+\frac{3}{D}-\frac{3}{2D^2}+\frac{1}{3D^3}\right].
\label{eq:fluxapp}
\end{eqnarray}
where $D=1+0.71$ GeV$^2$ $(\gamma_L^2/\omega^2)$.
One also alternatively estimate the photon flux from the relativistic point like
charge Z passing a target at a minimum impact parameter $b_{min}$:
\begin{eqnarray}
N_{\gamma/Z}(\omega)=\frac{2Z^2\alpha_{em}}{\pi}\left[\zeta K_0(\zeta)K_1(\zeta)-\frac{\zeta^2}{2}(K_1^2(\zeta)-K_0^2(\zeta))\right]\nonumber\\
\label{eq:fluxana}
\end{eqnarray}
where $K_0$ and $K_1$ are the modified Bessel functions of the second kind,
$\zeta=\omega b_{min}/\gamma_L$, where $b_{min}$ is the minimal admitted 
distance in the impact parameter space chosen to suppress the strong interaction
and $b_{min}=0.7$ fm for proton.  
In Fig.~1 we have compared the photon flux from the proton 
$N_{\gamma/p}$ for the 
$\Upsilon$ photoproduction,  $p Pb \rightarrow p+Pb+\Upsilon$,
from the exact expression Eqs.~\ref{eq:fluxexact}-\ref{eq:fluxsup} 
(presented by red curve)  and the  
approximate DZ expression Eq.~\ref{eq:fluxapp}
(presented by black curve). The blue curve (referred as FF)
is without SI suppression ($\Gamma_{pA}(b)=1$ in Eq.~\ref{eq:fluxexact}).
Upper panel shows photon flux for the 
$pPb$ collisions at $\sqrt s=$ 5.02 TeV, 
$Pb$ towards +ve rapidity with $E_{Pb}=1.58$ TeV and proton towards -ve rapidity with
$E_p=4 $ TeV, $Pb+p\rightarrow Pb+p+\Upsilon$. Lower panel of Fig.~1 
shows photon flux for $pPb$ collisions at $\sqrt s=$ 8.16 TeV, 
$Pb$ towards +ve rapidity with $E_{Pb}=2.56$ TeV and proton towards -ve rapidity with
$E_p=6.5 $ TeV.
 It is observed that,
the strong interaction between proton-nucleus
reduce the photon flux substantially 
at large -ve rapidities (high photon energies). 
\begin{figure}[h]
\includegraphics[width=75mm]{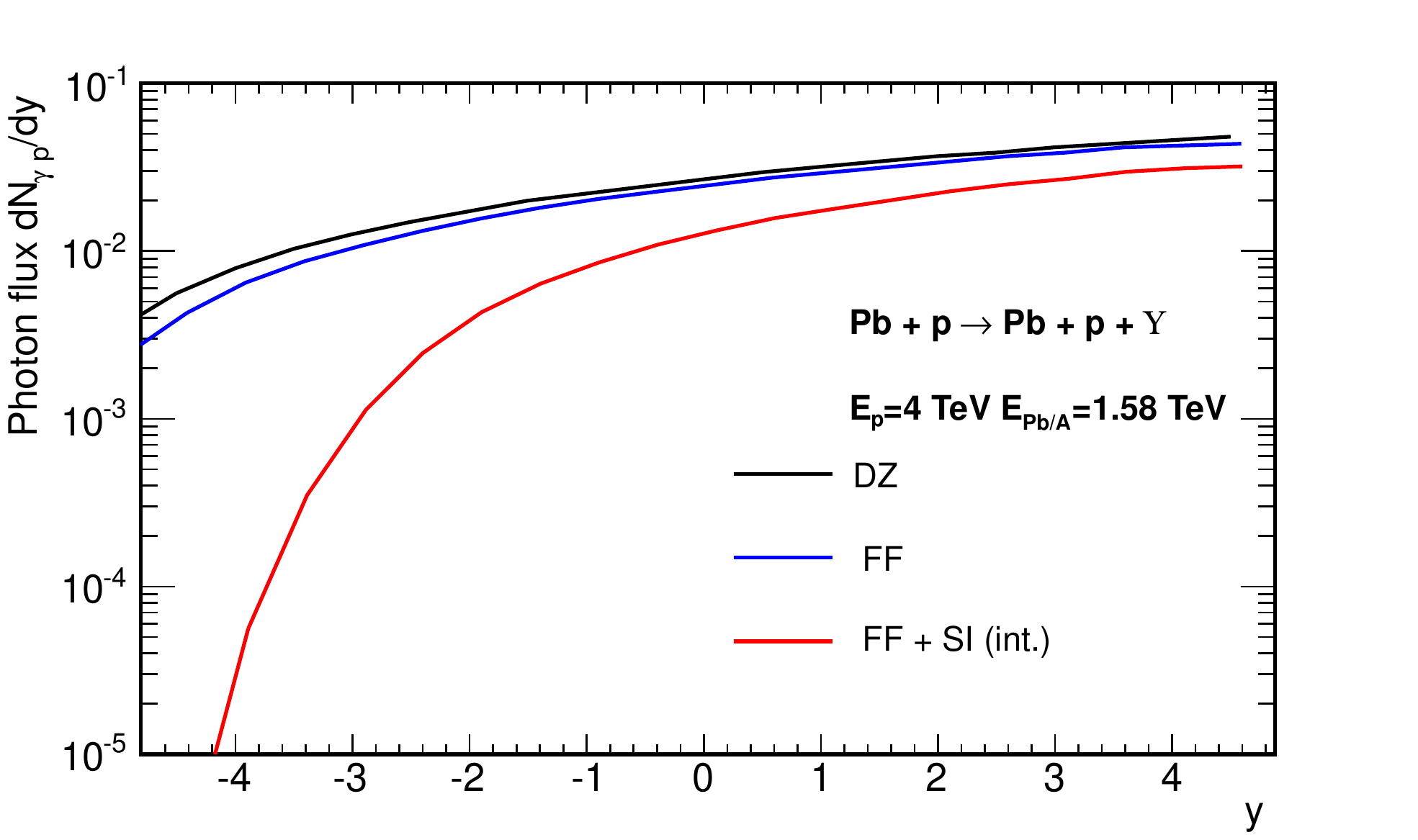}
\includegraphics[width=75mm]{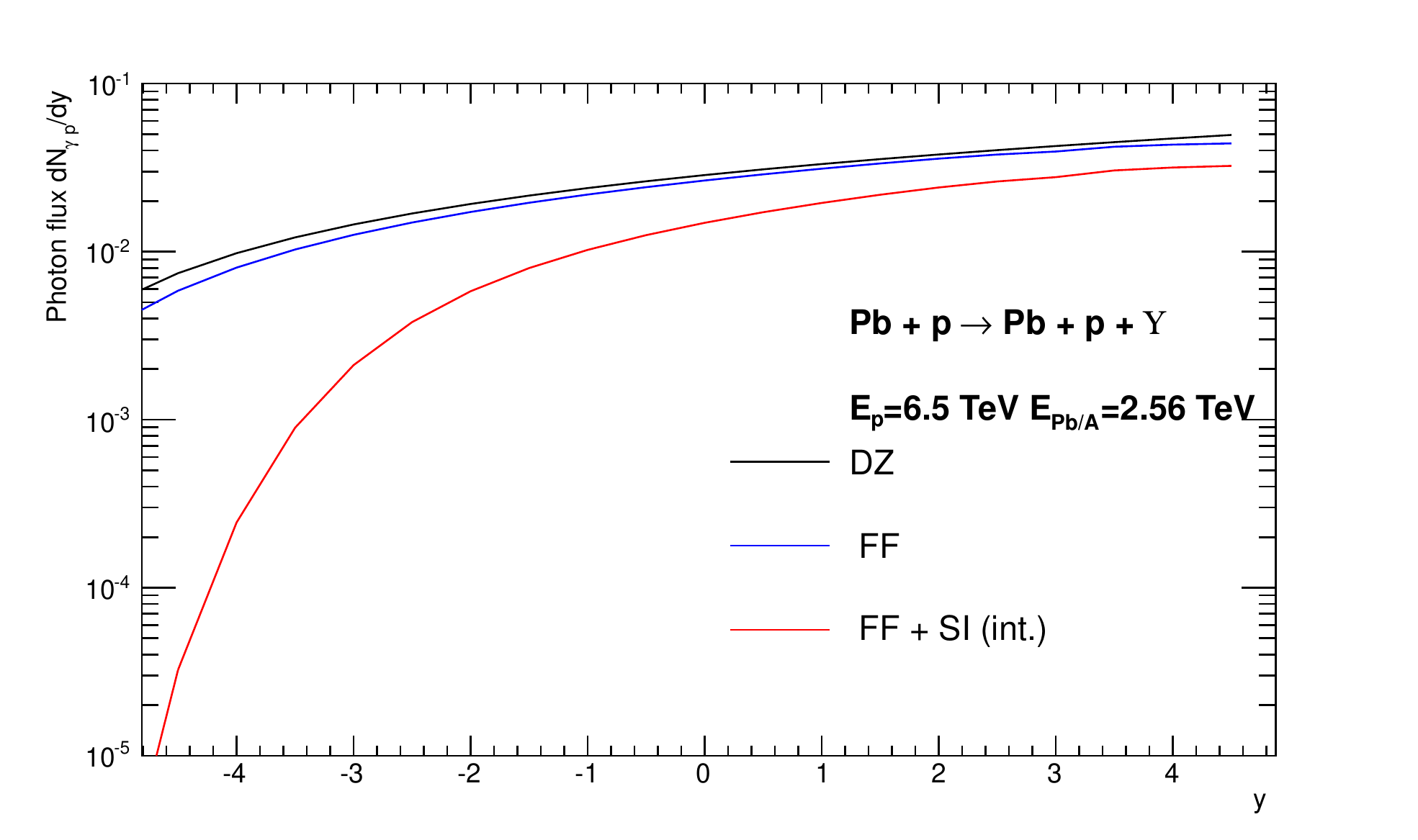}
\label{fig:photon_flux_pPb}
\vspace{-0.2cm}
\caption{The flux of photons from proton $N_{\gamma/p}$ as a function of  $\Upsilon$ (1S) rapidity $y$ 
in proton-Pb UPC. 
Upper panel shows photon flux for the 
$pPb$ collisions at $\sqrt s=$ 5.02 TeV, 
$Pb$ towards +ve rapidity with $E_{Pb}=1.58$ TeV and proton towards -ve rapidity with
$E_p=4 $ TeV, $Pb+p\rightarrow Pb+p+\Upsilon$.  Lower panel 
shows photon flux for $pPb$ collisions at $\sqrt s=$ 8.16 TeV, 
$Pb$ towards +ve rapidity with $E_{Pb}=2.56$ TeV and proton towards -ve rapidity with
$E_p=6.5 $ TeV.
DZ refers to Eq.~\ref{eq:fluxapp}, FF refers to
$\Gamma_{pA}(b)=1$ in Eq.~\ref{eq:fluxexact}.}
\end{figure}

Photon flux from nucleus can be similarly
be estimated by using Eqs.~\ref{eq:fluxexact}-\ref{eq:fluxsup}
with appropriate nuclear charge form factor in Eq.~\ref{eq:fluxim}. 
In this study, we preferred to use the point-like
expression Eq.~\ref{eq:fluxana} for nucleus, following previous studies
~\cite{guzey01,Guzey:2013aa,Guzey:2013bb,
Lappi:2013,Adeluyi:2012,Adeluyi:2013,sampaio,Goncalves:2014},
as the difference between the full calculation and
 point-like expression  was observed to be insignificant
for the photon flux from nucleus 
with $b_{min}=1.15 R_{Pb}$~\cite{guzey01,Guzey:2016}.
For PbPb collisions we used the point-like charge expression for the photon flux for nucleus  
Eq.~\ref{eq:fluxana} with minimum impact parameter $b_{min}=2R_{Pb}$.
  
\section{\label{sec:s3}Photoproduction of $\Upsilon$}
The rapidity distribution of vector meson  ($\Upsilon$) 
production in  proton-nucleus UPC interaction with
proton from the right and nucleus from the left,
is given by the sum of the two terms, 
each term in Weizsacker-Williams (WW) approximation~\cite{Weiz:1934,Will:1934}
is the product of photon flux and the cross-section of 
vector meson  photoproduction:
\begin{eqnarray}
\frac{\sigma_{AB\rightarrow AB \Upsilon} (y)}{dy}=N_{\gamma/A}(y)\sigma_{\gamma B \rightarrow
\Upsilon B} (y)&\nonumber\\ 
+ N_{\gamma/B}(-y)\sigma_{\gamma A \rightarrow \Upsilon A} (-y)&
\label{eqn1}
\end{eqnarray}
Here $N_{\gamma/A(B)}(y)$ is the photon flux of proton or nucleus;
$y=\ln(2\omega/M_{\Upsilon})$ is the rapidity of $\Upsilon$ 
where $\omega$ is the photon energy and 
$M_{\Upsilon}$ is the mass of the  $\Upsilon$.
The first term corresponds to photon from nucleus(A)
while the second term is due to photon flux from 
proton(B). As the photon flux $\propto Z^{2}$ and 
have support only small value of $\omega$, dying exponentially
at large value of $\omega$, the first term in r.h.s. ($\gamma p$ contribution)
dominates and peaks at positive rapidity while the second term
($\gamma A$ contribution)  peaks at negative rapidity. 

The cross section of exclusive elastic photo-production of 
$\Upsilon$ on $H$ ($H\equiv p,A$) can be written as
\begin{eqnarray}
\sigma_{\gamma H \rightarrow \Upsilon H} (y) =\frac{d\sigma_{\gamma H \rightarrow
\Upsilon H}}{dt}\arrowvert_{t=0}\int dt |F_{H}(t)|^2\nonumber\\
\end{eqnarray}
where $d\sigma_{\gamma H \rightarrow \Upsilon H}/dt|_{t=0}$ is the
forward scattering amplitude and $F_H(t)$ is the charge form factor of the hadron (nucleus). 
Using  leading order (LO)
approximation, the scattering amplitude for elastic photoproduction of $\Upsilon$
 from proton or a nucleus reads \cite{guzey01,Ryskin:2013jmr}: 
\begin{eqnarray}
\frac{d\sigma_{\gamma H \rightarrow \Upsilon H}(W_{\gamma p},t=0)}{dt}=\frac{M_{\Upsilon}^3 \Gamma_{ee} \pi^{3}}
{48 \alpha_{e.m.}\mu^8}&(1+\eta^{2}) \nonumber\\
~R_g^2 F^{2} (Q^{2})~[\alpha_s(Q^{2})\frac{xG_{H}(x,Q^{2})}{A}]^{2} & 
\label{eq:pht0}
\end{eqnarray}
where $\Gamma_{ee}$ is the width of $\Upsilon$ electronic decay; 
$\alpha_{em}$ is the fine structure constant; $\alpha_s(Q^2)$ is the running
strong coupling constant; $x=M_\Upsilon^2/W_{\gamma p}^2$, is the fraction of nucleon momentum
carried by nucleons, $W_{\gamma p}$
is the $\gamma p $ center of mass energy;
$G_H(x,Q^2)$ is the  gluon
distribution
in the proton (nucleus)  evaluated at momentum transfer
$Q^2=(M_{\Upsilon}/2)^2$.  The relevant $x$ region in CMS  
is $\approx 10^{-2} -10^{-4}$  at central rapidities ($|y|<2.4$).   
The factors $(1+\eta^2)$, $R_g^2$ and  $F^2(Q^2)$ corresponds
to correction due to real part,
skewness and  next-to leading (NLO), respectively. 

The $t$ or the momentum-squared transferred
dependence of the cross section for
the proton target is generally parameterized
in the form of rapidly decreasing exponential function, $e^{-B(W_{\gamma p})|t|}$,
where  the slope parameter $B(W_{\gamma p})$  depends weekly on energy.
Here we use the  Regge motivated expression for the slope parameter which 
is obtained from the fitting the data of exclusive J/$\psi$ photoproduction~\cite{Ryskin:2013jmr}, 
$B_{\Upsilon}(W_{\gamma p})=4.63+4.0\times 0.06~\ln(W_{\gamma p}/90~\mbox{GeV})$. 
Hence for proton, the photoproduction of $\Upsilon$ reads,
\begin{eqnarray}
\sigma_{\gamma p \rightarrow \Upsilon p} (W_{\gamma p})=\frac{1}{B(W_{\gamma p})} \frac{d\sigma_{\gamma p \rightarrow \Upsilon p}}{dt} \arrowvert_{t=0}.
\label{eq:pht1}
\end{eqnarray}

To evaluate the factors $\eta$ and $R_g$,
we have fitted the shape of each gluon distribution $x G_{T}(x,Q^2) \propto 1/x^{\lambda}$
(details of gluon distributions used is given in next section) 
and determined the corresponding $\lambda$ from the fit.
The $\lambda$ factors  for $Q^2=22.4$ GeV$^2$ (corresponds to $\Upsilon$(1S))
are given in Table~\ref{table:norm}. 
The $\lambda$ factors for CTEQ6L1(CTEQ6M) for $Q^2=25.1$ GeV$^2$ and $Q^2=26.8$ GeV$^2$
are $0.39(0.40)$ and $0.29(0.29)$ respectively. The factor $\eta$ is evaluated
using Gribov-Migdal relation~\cite{Gribov:1969,Ryskin:2008rt},
 $\eta=\tan(\pi \lambda /2)$.
In collinear factorization for hard exclusive process, one should use
gluon generalized parton distribution (GPD) and phenomenologically
it is being treated by introducing enhancement factor $R_g$~\cite{guzey01}
given by,
\begin{eqnarray}
R_g&=&\frac{2^{3+2\lambda}}{\sqrt{\pi}}\frac{\Gamma(\frac{5}{2}+\lambda}{\Gamma(4+\lambda)}
\end{eqnarray}
\begin{table}[!htb]
\vspace{-0.4cm}
\caption{\label{tab:norm}
The fitting factor for gluon PDFs $\lambda$.}
\label{table:norm}
\begin{ruledtabular}
\begin{tabular}{|c|c|c|c|c|c|}
Param.&MSTW08&CTEQ6L&CTEQ6L1&JMRTLO&CTEQ6M\\
\hline
$\lambda$&0.41&0.36&0.38&0.36&0.28\\
\hline
\end{tabular}
\end{ruledtabular}
\end{table}
The suppression factor $F^2(Q^2)$ contains all effects beyond
the leading order collinear factorization  and discussed
in Ref.~\cite{Frankfurt:2001, Ryskin97}.
However this correction mainly 
influence the absolute normalization, but not the $x$ dependance
of the cross-section.  We have used $F^2(Q^2)=1$  
and $Q^2=M_{\Upsilon}^{2}/4$,
which is the  non-relativistic approximation
that neglects the transverse momenta
of b-quarks in the $\Upsilon$ wave function. 
\begin{figure}[!htb]
\includegraphics[width=75mm]{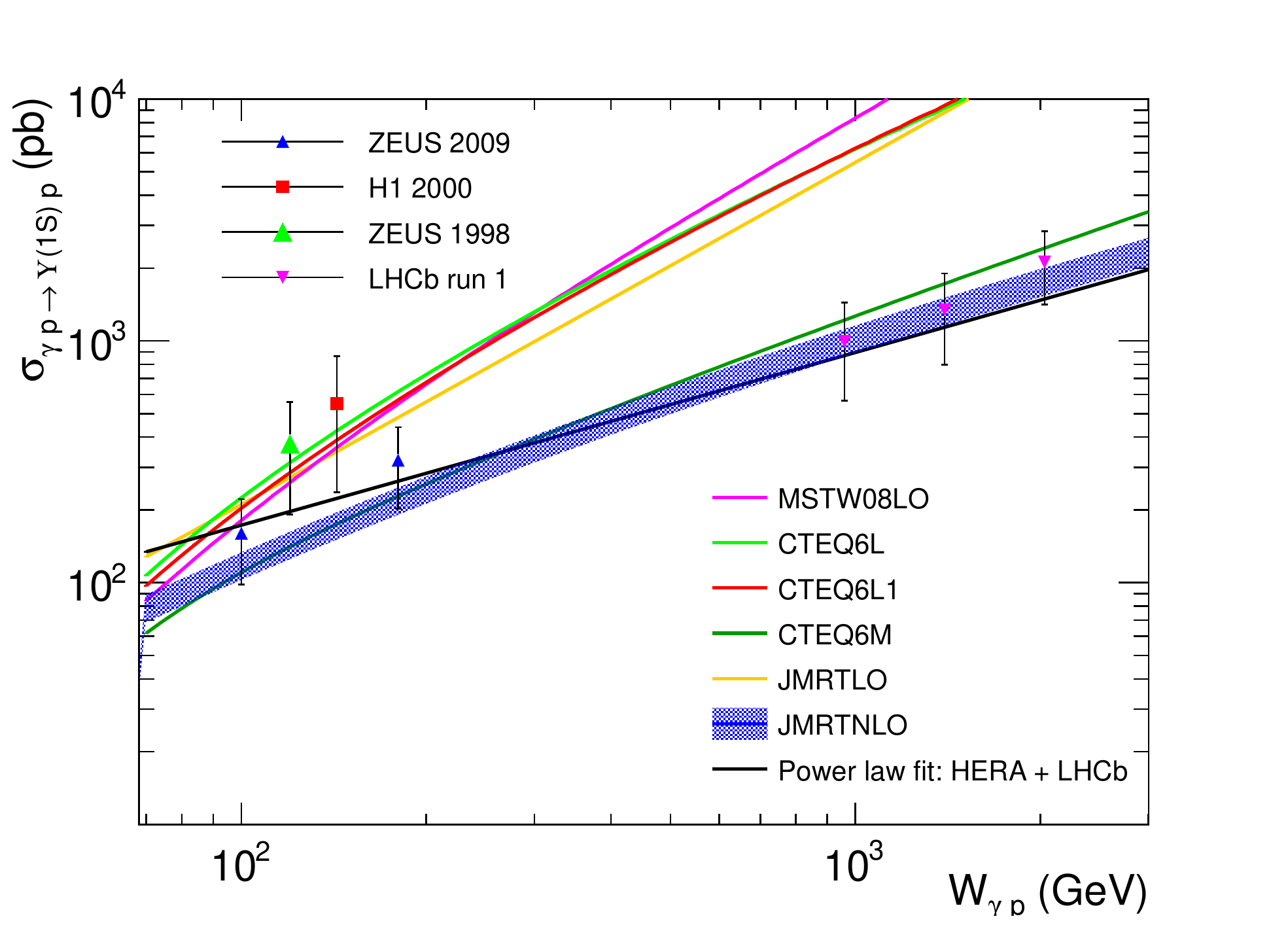}
\label{fig:pht1}
\vspace{-0.2cm}
\caption{Photoproduction cross section of $\Upsilon$ (1S),
 $\sigma_{\gamma p \rightarrow \Upsilon(1S) p}$ with  photon-proton
center of mass energy $W_{\gamma p}$ compared with 
the experimental data from HERA~\cite{herazeus,ZEUS:2009,ZEUS:2012} and  
LHCb~\cite{LHCb:2015}. pQCD prediction with
different gluon distribution compared with power law "Fit" to HERA + LHCb data.}
\end{figure}
\subsection{\label{sec:sub1}Photoproduction of $\Upsilon$  from proton}
Different parameterization of the gluon distribution
in the proton are being used to estimate the
photoproduction from  $\sigma_{\gamma p \rightarrow \Upsilon(1S) p}$
and the results are shown in the Fig.~2.
The LO pQCD estimation are being compared with HERA, LHCb  data
of exclusive photoproduction cross section of $\Upsilon(1S)$ with $W_{\gamma p}$.
It also shows the power law fit to the  data on 
$d\sigma_{\gamma p \rightarrow \Upsilon(1S) p}(W_{\gamma p},t=0)/dt$
of HERA + LHCb and multiplied by the $B_{\Upsilon}(W_{\gamma p})$ using Eq.~\ref{eq:pht1} 
to get the photoproduction cross section $\sigma_{\gamma p \rightarrow \Upsilon(1S) p}$. 
Gluon parameterization used are MSTW08~\cite{MSTW08}, CTEQ6L~\cite{CTEQ6L},
CTEQ6L1~\cite{CTEQ6L}, JMRTLO~\cite{Ryskin:2013jmr}, CTEQ6M~\cite{CTEQ6L} and  
JMRTNLO~\cite{Ryskin:2013jmr}. 
Fig.~2 shows the
LO pQCD predictions from Eqs.~\ref{eq:pht0}-\ref{eq:pht1} 
at $Q^2=M_{\Upsilon(1S)}^2/4$ for different
gluon parameterization as discussed above.    
We have shown JMRTLO and JMRTNLO~\cite{Ryskin:2013jmr} cross-section of 
$\Upsilon(1S)$ photoproduction with $W_{\gamma p}$, for comparison
and could not accomodate the error band for LO.
The power law fit to
HERA+LHCb data with $\delta=0.76\pm 0.27$ 
($\sigma_{\gamma p \rightarrow \Upsilon (1S) p} \propto 
1/B_{\Upsilon (1S) (W_{\gamma p})} \times W_{\gamma p}^{\delta}$),
 shown by black curve, is less steep than  LO predictions
and comparable to  NLO gluon PDFs, CTEQ6M and JMRTNLO
predictions, the power law fit to which approximately gives 
$\delta=0.84$.  

\subsection{\label{sec:sub2}Photoproduction of $\Upsilon$  from nucleus}
In case of nuclear target, the photoproduction  is given by, 
\begin{eqnarray}
\sigma_{\gamma A \rightarrow \Upsilon A}(W_{\gamma p})=&S^{2}_{A}(W_{\gamma p})\frac{d\sigma_{\gamma p \rightarrow \Upsilon p}}{dt}\arrowvert_{t=0}\nonumber\\
&\times\Phi_A(t_{\mbox{min}})  
\label{eq:pht2}
\end{eqnarray}
where 
\begin{eqnarray}
\Phi_A(t_{\mbox {min}})=\int_{t_{\mbox {min}}}^{\infty} dt |F_{A}(t)|^2
\label{eq:pht3}
\end{eqnarray}
and $t_{\mbox {min}}=-M_{\Upsilon}^4 m_{N}^2/W_{\gamma p}^4$
is the minimal momentum transfer to the nucleus; 
$F_{A}(t)$ is the nuclear form factor which is given by the Fourier transform
of the nuclear density distribution $F_A(t)=\int d^3 r \rho(r) e^{i{\bf q}\cdot{\bf r}}$, where ${\bf q}$ is the momentum transferred, $\rho(r)$
is the nuclear density  approximated as modified
hard sphere~\cite{Adeluyi:2011}.
$S_A(W_{\gamma p})$ is the nuclear suppression factor given by,
\begin{eqnarray}
S_A(W_{\gamma p})&=&\frac{G_A(x,Q^{2})}{AG_N(x,Q^{2})}\times \left[\frac{(1+\eta_{A}^{2})R_{g,A}^2}{(1+\eta^{2})R_g^2}\right]^{1/2}\nonumber\\
&=&R(x,Q^{2})\times \kappa_{A/N}. 
\label{eq:pht4}
\end{eqnarray}
Here, $G_N(x,Q^{2})$ and  $G_A(x,Q^{2})$ are the gluon density of
nucleon and nucleus respectively; 
$R(x,Q^{2})$ is the nuclear gluon modification factor.
Due to nuclear gluon shadowing at small values of
$x$, $G_A(x,Q^{2}) < AG_N(x,Q^{2})$
and correspondingly $R(x,Q^{2})<1$. In addition, due to different growth of the 
nuclear gluon density $G_A(x,Q^{2})$ with $x$ than free proton, 
$G_A(x,Q^{2})\propto 1/x^{\lambda_{A}}$ where $\lambda_A < \lambda_p$
($\lambda_p$ corresponds to proton),
results to the $\kappa_{A/N}$ factor ~\cite{guzey01}.
\begin{figure}[!ht]
\includegraphics[width=85mm]{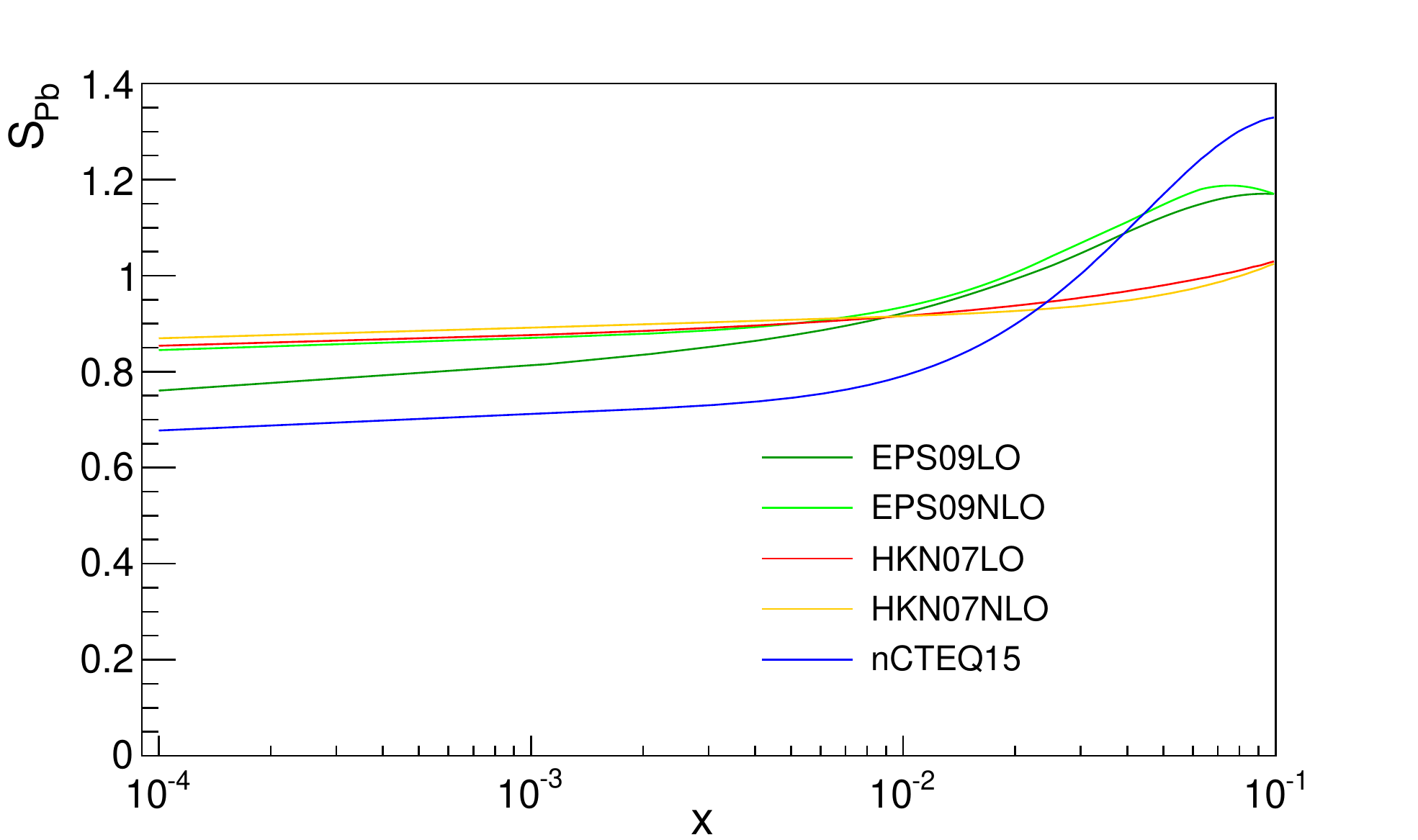}
\label{fig:pht1}
\vspace{-0.2cm}
\caption{Nuclear suppression factor of EPS09~\cite{Eskola:2009}
 gluon distribution for  $\Upsilon$ (1S),
as defined in  Eq.~\ref{eq:pht4}.}
\end{figure}

We have used  gluon shadowing parameters from 
three recently available nuclear parton distribution functions:
(i)  EPS09  nuclear PDF at leading order and next-to-leading order~\cite{Eskola:2009}
(ii)  HKN07 at LO and NLO~\cite{HKN07:2007}
(iii) nCTEQ15~\cite{nCTEQ15:2015}
and estimated the suppression factor $S(W_{\gamma p})$ for
for $\Upsilon(1S,2S,3S)$  ($Q^2=22.4, 25.1, 26.8$ GeV$^2$ for 1S, 2S and 3S respectively) 
in the low $x$ region accessible in LHC experiments.
We have used $\kappa_{A/N}=0.87$ neglecting its variation with different gluon distributions
~\cite{guzey01}.
Fig.~3 shows the variation of suppression factor of
 Pb nucleus $S_{Pb}$  with $x$ for 
  $\Upsilon$(1S) with three nuclear shadowing parameterizations.
 We should mention here that EPS09LO use CTEQ6L1 
 gluon distribution for free proton whereas EPS09NLO use CTEQ6M gluon PDF for proton. 
Hence results with CTEQ6L1(CTEQ6M) for $\gamma p$ and EPS09LO(EPS09NLO) for $\gamma Pb$,
corresponds to the right combination of gluon distribution which  are being compared with
other combinations of gluon distributions for proton and nucleus.

\section{\label{sec:s4}Photoproduction cross section with rapidity}
We now present the results of photoproduction of $\Upsilon$(1S)
in the framework of leading-order QCD with two gluon exchange.
Three factors,  square of gluon distribution, photon flux and integrated
nuclear form factor, determines the rapidity distribution of cross section.   
\subsection{\label{sec:sub3}Cross section for pPb collisions}
\begin{figure}[!ht]
\includegraphics[width=85mm]{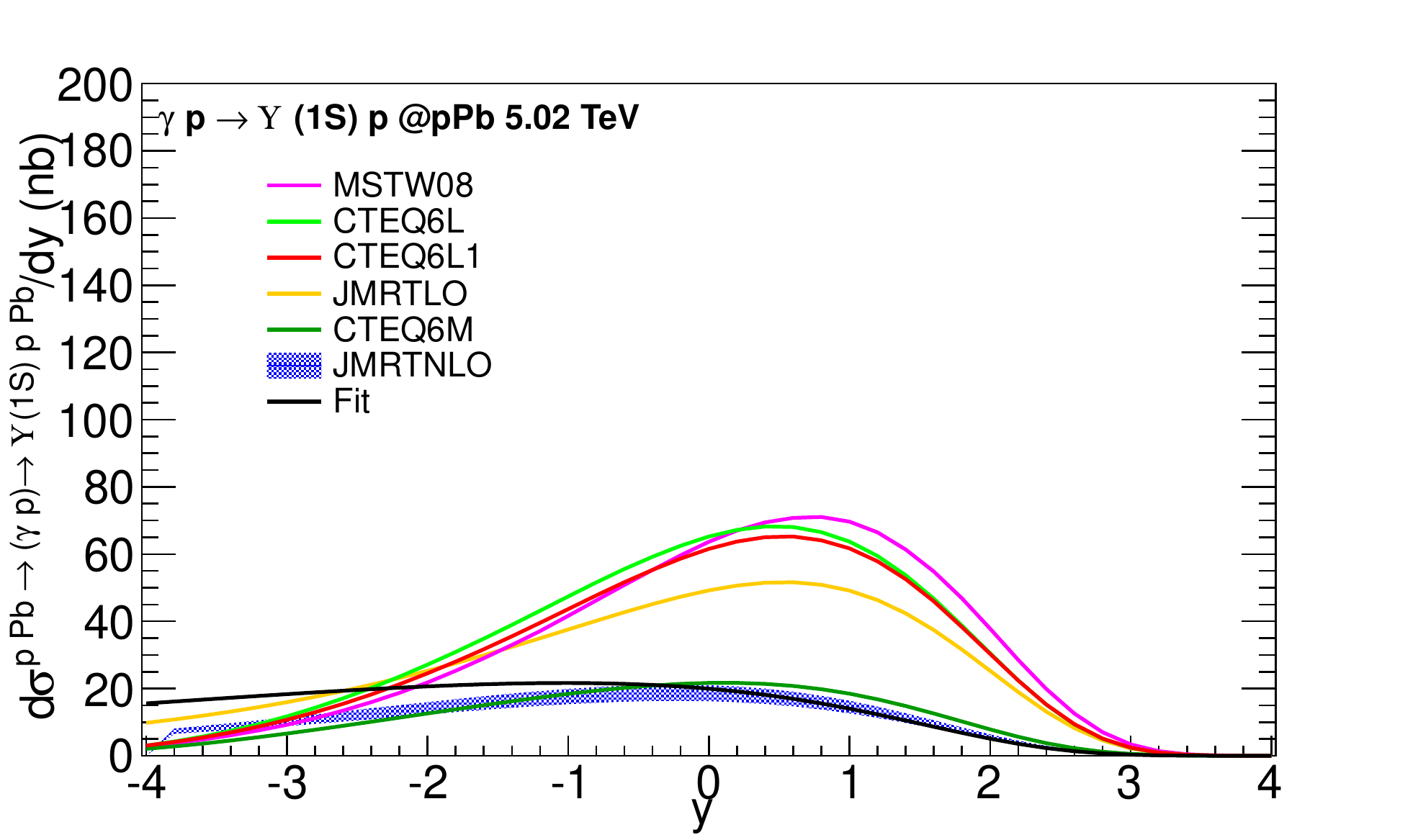}
\hspace{-9.8mm}\llap{\raisebox{20mm}{\includegraphics[height=25mm]{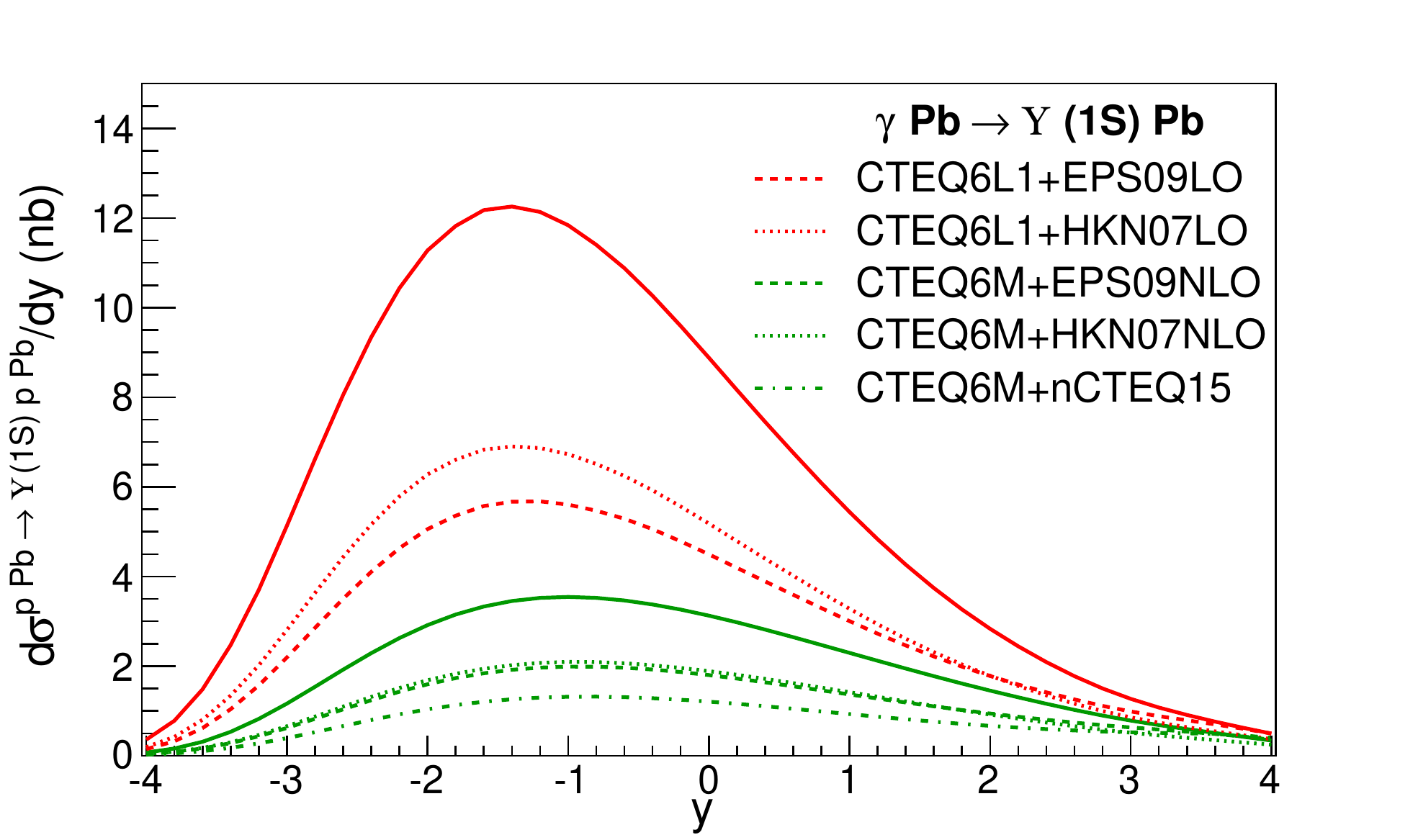}}}
\label{fig:fig3}
\vspace{-0.2cm}
\caption{The rapidity distribution of $\Upsilon$ (1S)
photoproduction cross section for pPb collisions at $\sqrt s= 5.02$ TeV.
 The main and
insert figure shows $\gamma p$ and $\gamma Pb$ contributions respectively.}
\end{figure}
Fig.~4 shows the  rapidity distribution of the
$\Upsilon$(1S) photoproduction in
proton-Pb UPC collisions integrated over momentum transfer $t$
in the LHC kinematics as estimated by Eq.~\ref{eqn1}.
 Photon flux includes
the effect of strong interaction suppression 
as discussed in Sections~\ref{sec:s2}.
Here we used appropriate
$t$ dependence of cross section for proton and nucleus    
as discussed in Sections~\ref{sec:sub1} and ~\ref{sec:sub2} respectively.
Figure shows the estimations using MSTW08 (magenta)~\cite{MSTW08}, CTEQ6L (light green)~\cite{CTEQ6L},
CTEQ6L1 (red)~\cite{CTEQ6L}, CTEQ6M (dark green)~\cite{CTEQ6L}, JMRTLO (orange)~\cite{Ryskin:2013jmr}, JMRTNLO (blue)~\cite{Ryskin:2013}  
gluon distribution functions and power law fit to HERA+LHCb data (black).
Fig.~4, main figure, shows the $d\sigma/dy$  distribution for 
$p Pb\rightarrow (\gamma p)\rightarrow \Upsilon (1S) p Pb$, the dominant contribution,
whereas insert figure shows the $\gamma Pb$ contribution. 
Insert figure  also shows the cross-section with different
nuclear gluon shadowing parameterization (EPS09~\cite{Eskola:2009}, HKN07O~\cite{HKN07:2007} 
and nCTEQ15~\cite{nCTEQ15:2015}) for CTEQ6L1 and CTEQ6M gluon PDF of free proton.
Solid lines show cross-section without shadowing
 whereas dashed curves are with nuclear gluon shadowing.
HERA + LHCb power law fit (black), gives comparable description as of NLO gluon PDF, CTEQ6M.

\begin{table}[!hb]
\vspace{-0.4cm}
\caption{
Cross section of photoproduction of $\Upsilon$(1S) 
at $\sqrt s_{NN}= 5.02$ TeV  and at $\sqrt s_{NN}= 8.16$ TeV in pPb collisions in CMS acceptance
$-2.4<y<2.4$ with different gluon PDF parameterizations.
}
\label{table:crosspPb}
\begin{ruledtabular}
\begin{tabular}{|c|c|c|c|c|c|c|}
Param.&
\multicolumn{3}{c|}{${\sqrt s}$=5.02 TeV} &\multicolumn{3}{c|}{${\sqrt s}$=8.16 TeV}\\
&\multicolumn{3}{c|}{$\sigma$(nb)} &\multicolumn{3}{c|}{$\sigma$(nb)}\\
\cline{2-7}
&{$\gamma$ p}&{$\gamma$ Pb} &Total &{$\gamma$ p}&{$\gamma$ Pb} & Total \\
\hline
MSTW08&229&42&271 &488&82&570\\
\hline
CTEQ6L&229&40&269 &451&73&524\\
\hline
CTEQ6L1&217&39&255 &434&71&505\\
\hline
JMRTLO&182&32&214&360&59&419\\
\hline
CTEQ6M&77&13&90 &127&19&146\\
\hline
JMRTNLO&69&11&80 &115&17&132\\
\hline
Fit&85&14&99&122&18&140\\
\hline
\end{tabular}
\end{ruledtabular}
\end{table}

\begin{figure}[!ht]
\includegraphics[width=85mm]{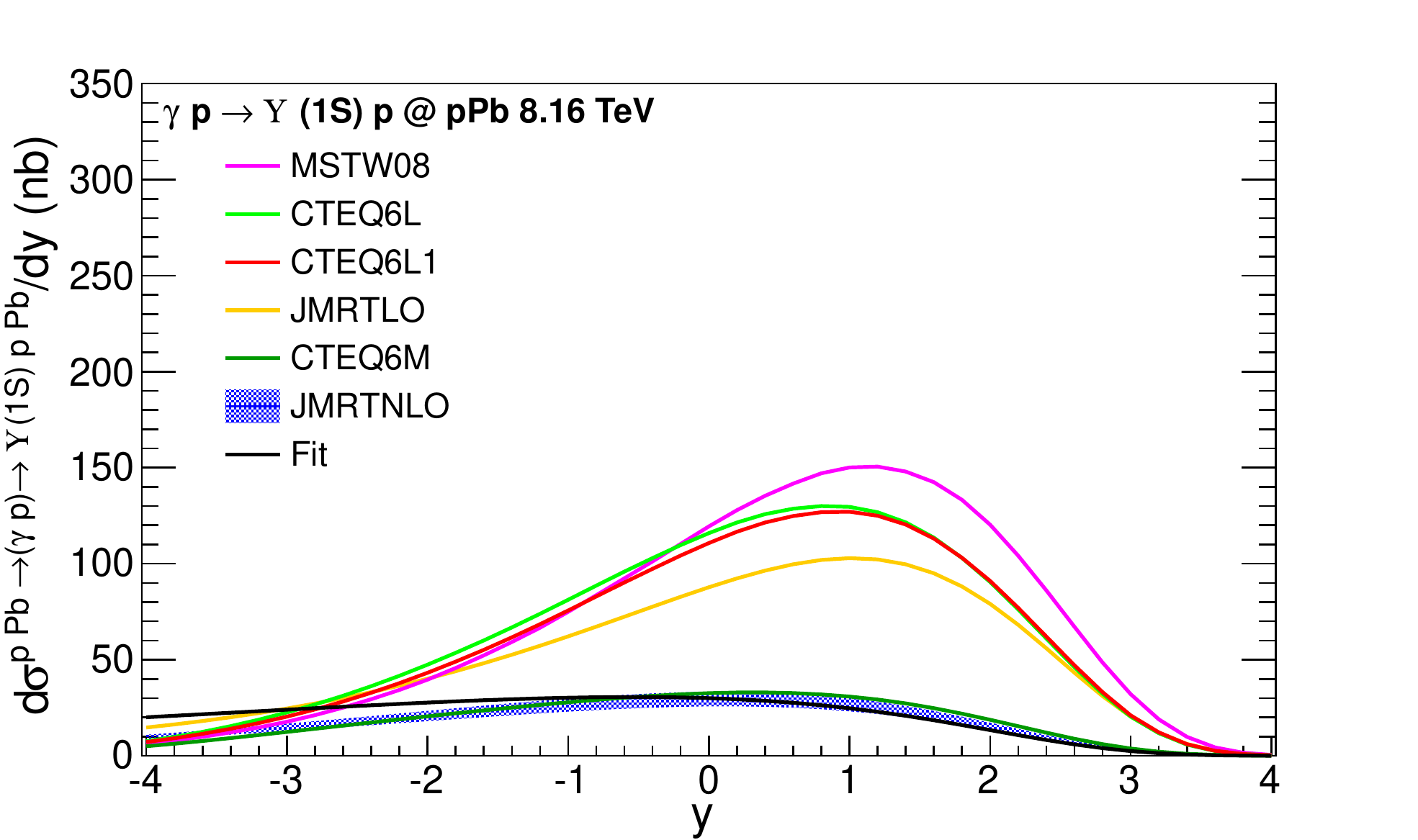}
\hspace{-9.8mm}\llap{\raisebox{23mm}{\includegraphics[height=22mm]{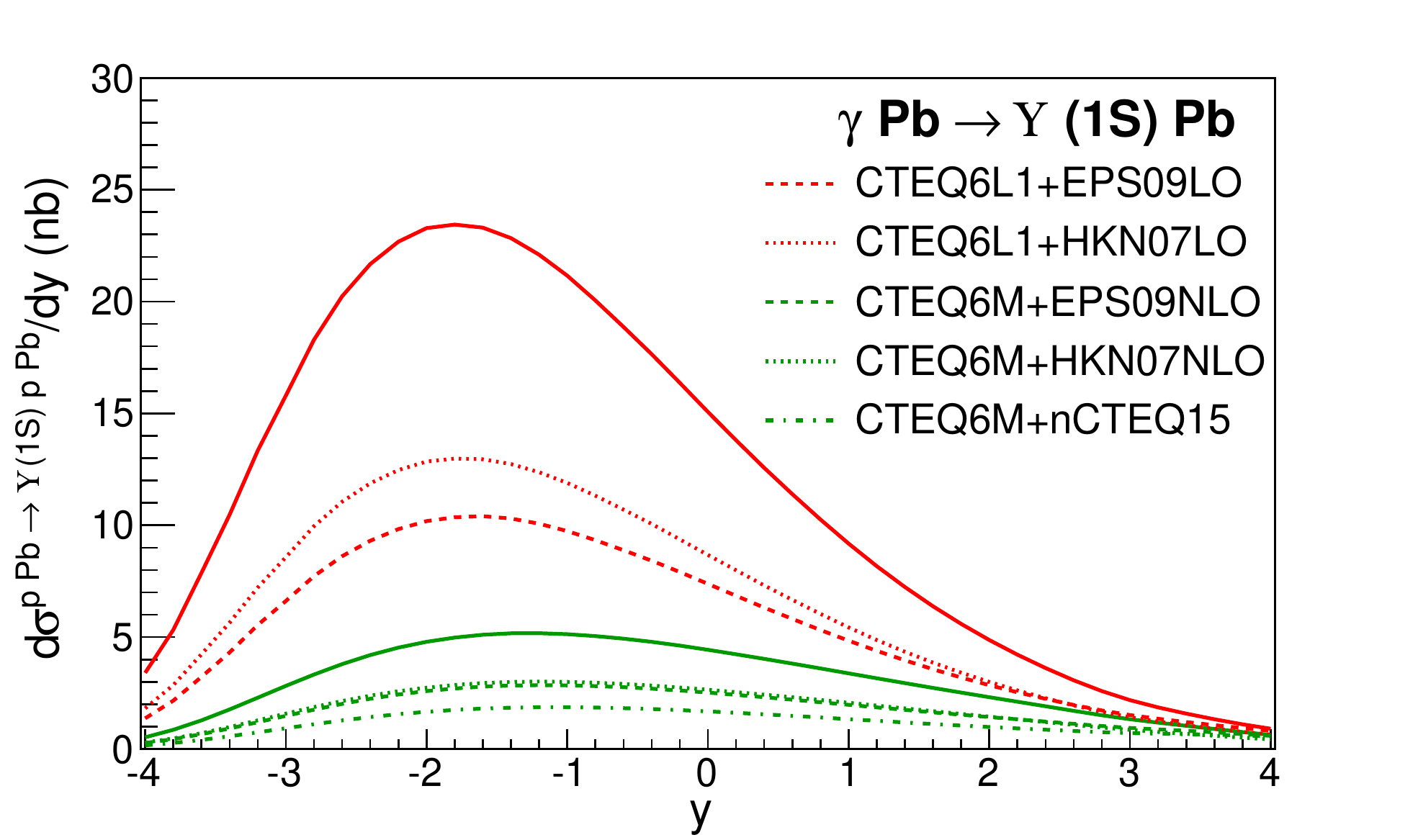}}}
\label{fig:fig5}
\vspace{-0.2cm}
\caption{The rapidity distribution of $\Upsilon$ (1S)
photoproduction cross section for pPb collisions at $\sqrt s= 8.16$ TeV. 
The main and
insert figure shows $\gamma p$ and $\gamma Pb$ contributions respectively.}
\end{figure}

We also estimate the $\Upsilon$(1S) photoproduction cross-section
for pPb collisions at $\sqrt{s_{NN}}=8.16$ TeV, Run2 LHC collision scenario.
Fig.~5 shows the  $d\sigma/dy$ distribution for $\Upsilon$ (1S)
 for ${\sqrt s}=8.16$ TeV. Similar to Fig.~4, main and insert figure of Fig.~5, shows
$\gamma p$ and $\gamma Pb$ contributions respectively.
Table~\ref{table:crosspPb} gives the rapidity integrated cross section of $\Upsilon$ (1S) in
CMS acceptance, i.e. $-2.4<y<2.4$  for ${\sqrt s}=5.02$ TeV and ${\sqrt s}=8.16$ TeV for different gluon PDF.
We here present, the $\gamma p$ and  $\gamma Pb$ contributions to the $\Upsilon$ (1S)
cross-section separately for different gluon PDF.

\begin{figure}[!ht]
\includegraphics[width=40mm]{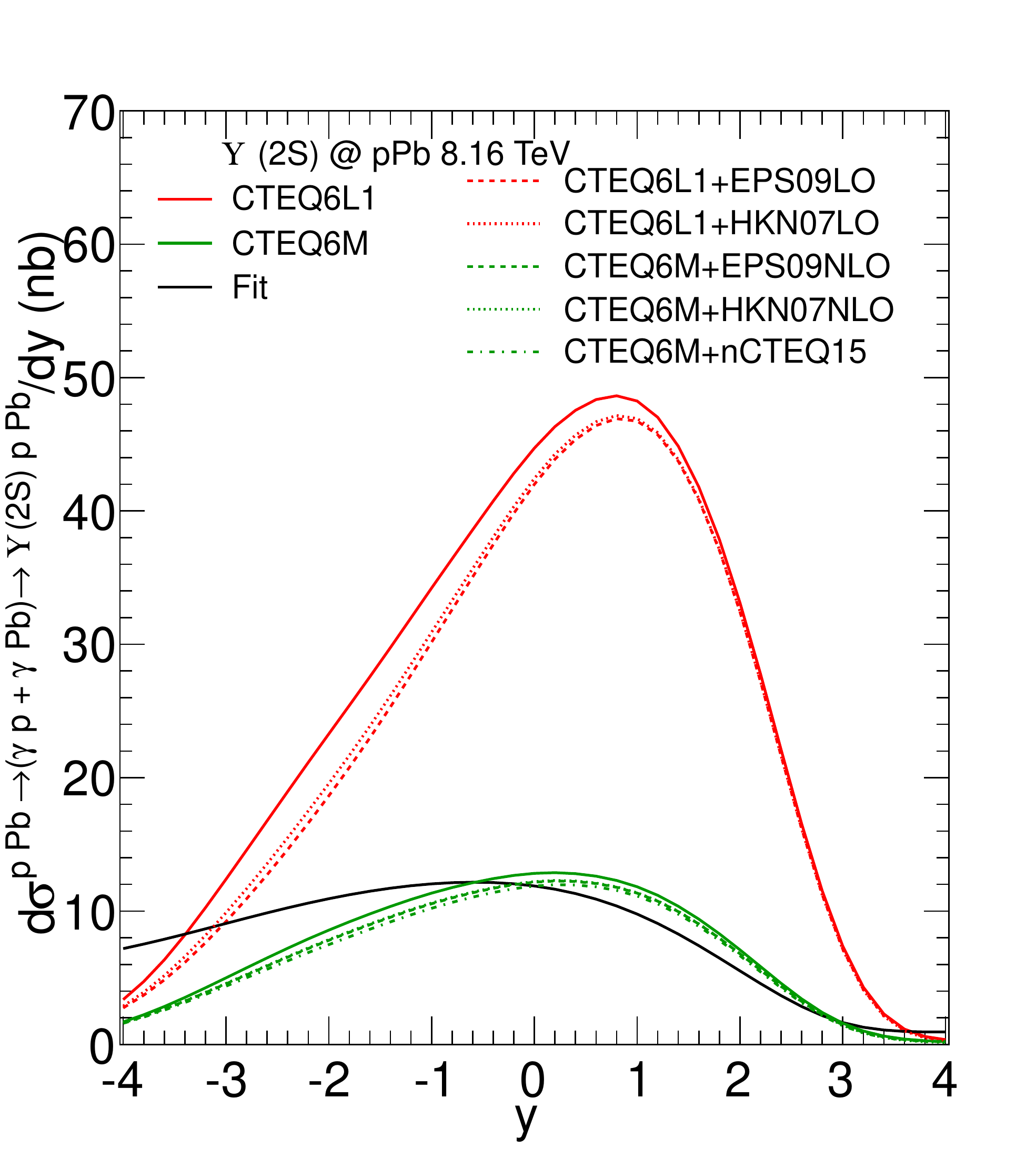}
\includegraphics[width=40mm]{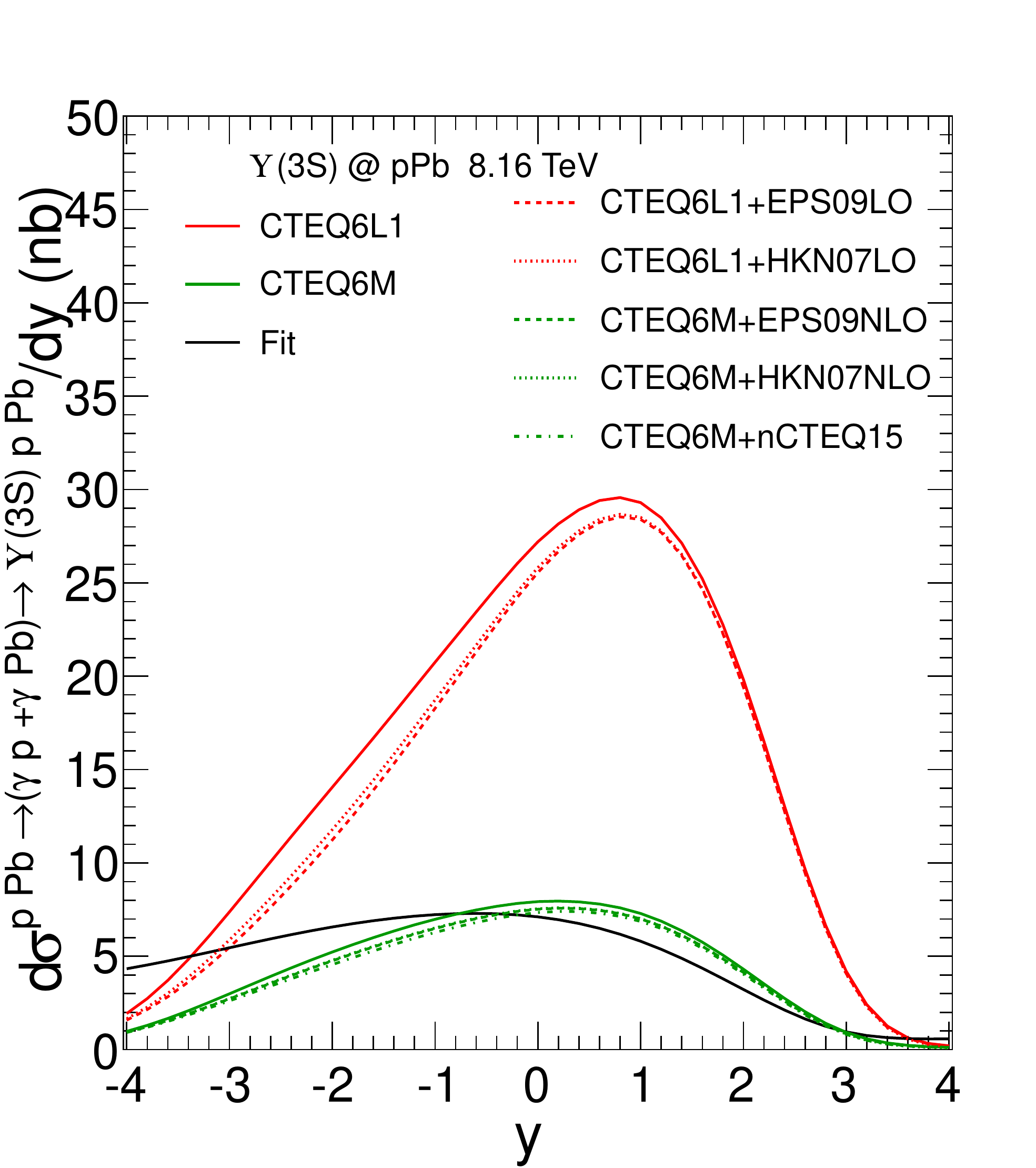}
\label{fig:fig7}
\vspace{-0.2cm}
\caption{The rapidity distribution of $\Upsilon$ (2S) (left plot)
and $\Upsilon$ (3S) (right plot) 
 photoproduction cross section for pPb collisions at $\sqrt s= 8.16$ TeV
 with different nuclear gluon shadowing parameterizations.}
\end{figure}
We also estimate the cross-section for $\Upsilon$ (2S) and  $\Upsilon$ (3S) in
 pPb collisions at $\sqrt s= 8.16$ TeV.
Fig.~6 shows the  $d\sigma/dy$ distribution for $\Upsilon$ (2S) (left plot)
and $\Upsilon$ (3S) (right plot) of total cross-section with different gluon
shadowing parameterizations for CTEQ6L1 and CTEQ6M gluon PDF.
\begin{table}[!ht]
\vspace{-0.4cm}
\caption{
Cross section of photoproduction of $\Upsilon$(1S) in pPb collisions 
at $\sqrt s= 5.02$ TeV  and $\Upsilon$(1S,2S,3S) in $\sqrt s= 8.16$ TeV in CMS acceptance
$-2.4<y<2.4$ and also in full acceptance (numbers in bracket) 
with different nuclear gluon shadowing parameterizations.
We also present the ratio of cross-section of $\sigma^{\Upsilon(2S)}$/$\sigma^{\Upsilon(1S)}$ 
and $\sigma^{\Upsilon(3S)}$/$\sigma^{\Upsilon(1S)}$.}
\label{table:crosspPbshad}
\begin{ruledtabular}
\begin{tabular}{|c|c|c|c|c|c|c|}
Param.&
5.02 TeV &\multicolumn{5}{c|}{8.16 TeV}\\
\cline{2-7}
&$\sigma$(nb) &\multicolumn{3}{c|}{$\sigma$(nb)} &\multicolumn{2}{c|}{Ratio}\\
\cline{2-7}
&$\Upsilon$ (1S) &$\Upsilon$ (1S) &$\Upsilon$ (2S) &$\Upsilon$ (3S)&2S/1S &3S/1S \\ 
\hline
CTEQ6L1&255&505&178&108&0.35&0.21\\
&(284)&(589)&(205)&(124)&(0.35)&(0.21)\\
\cline{2-7}
EPS09LO&236&468&164&100&0.35&0.21\\
&(260)&(538)&(188)&(113)&(0.35)&(0.21)\\
\cline{2-7}
HKN07LO&239&474&167&101&0.35&0.21\\
&(264)&(547)&(191)&(115)&(0.35)&(0.21)\\
\hline
CTEQ6M&90&146&50&31&0.34&0.21\\
&(103)&(175)&(59)&(36)&(0.34)&(0.21)\\
\cline{2-7}
EPS09NLO&84&138&47&29&0.34&0.21\\
&(96)&(164)&(56)&(34)&(0.34)&(0.21)\\
\cline{2-7}
HKN07NLO&85&138&47&29&0.34&0.21\\
&(97)&(164)&(56)&(34)&(0.34)&(0.21)\\
\cline{2-7}
nCTEQ15&82&134&46&28&0.34&0.21\\
&(94)&(159)&(54)&(33)&(0.34)&(0.21)\\
\hline
\end{tabular}
\end{ruledtabular}
\end{table}
Table~\ref{table:crosspPbshad} shows the cross-section
for $\Upsilon$ (1S) at ${\sqrt s}=5.02$ TeV and $\Upsilon$ (1S, 2S, 3S)
at ${\sqrt s}=8.16$ TeV pPb collisions with different nuclear gluon shadowing.  
Integrated cross-section for CMS acceptance and
in full acceptance (numbers in bracket)
with different nuclear gluon shadowing parameterizations are presented.
Due to large photon flux from Pb nucleus,
$\gamma p$ is the dominant part of the cross section,
the $\gamma Pb$ contribution is small and thus
 nuclear shadowing does not affect significantly the
 total $d\sigma/dy$ distribution as well as total integrated cross-section. 
We also estimate the ratio of cross-sections 
$\sigma^{\Upsilon(2S)}/\sigma^{\Upsilon(1S)}$
and $\sigma^{\Upsilon(3S)}/\sigma^{\Upsilon(1S)}$
for different gluon PDF in the rightmost two columns
 of Table~\ref{table:crosspPbshad}.

\subsection{\label{sec:s4}Cross section for PbPb collisions}
\begin{figure}[!hb]
\includegraphics[width=85mm]{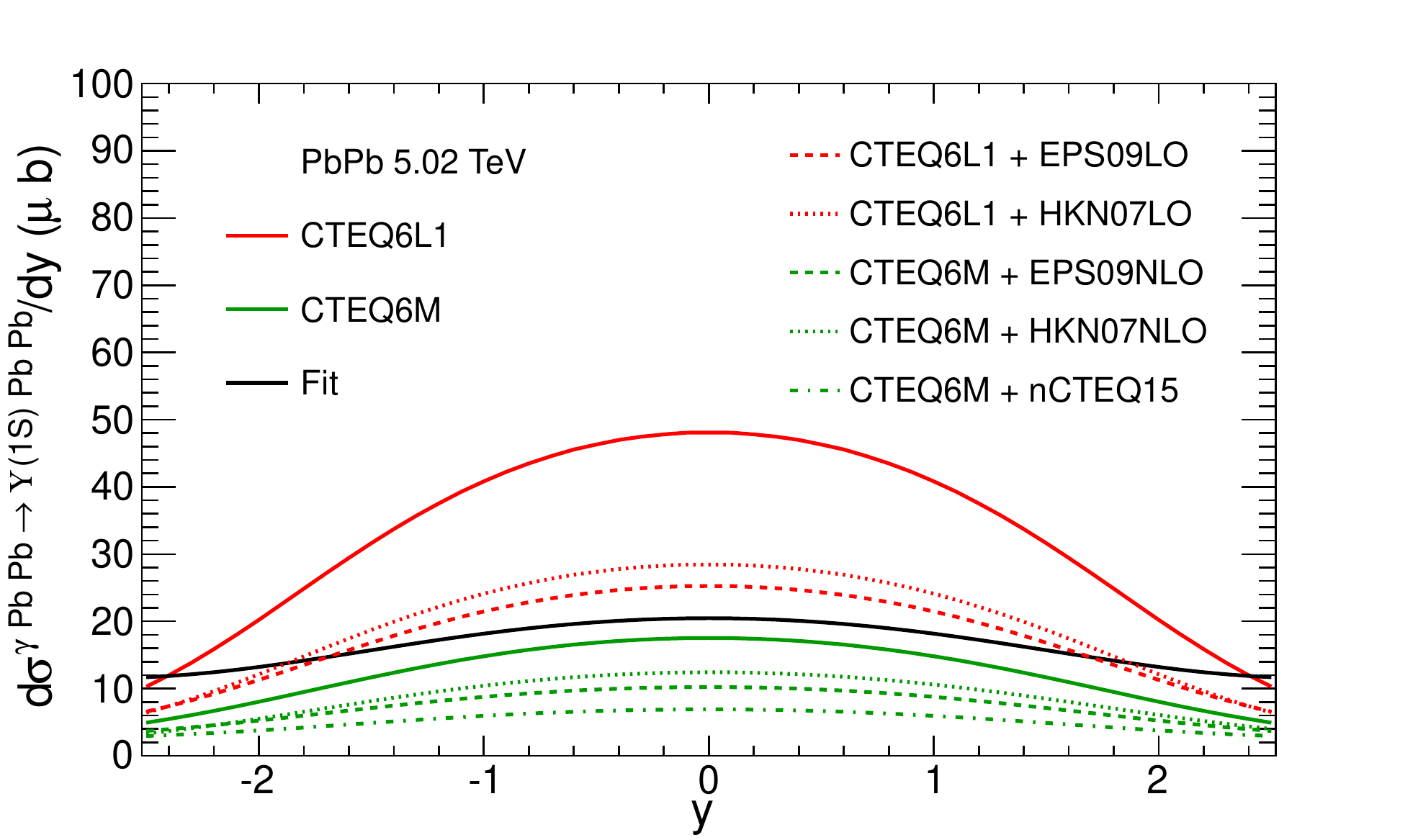}
\label{fig:fig6}
\vspace{-0.2cm}
\caption{The rapidity distribution of $\Upsilon$ (1S)
photoproduction cross section for $PbPb$ collisions at ${\sqrt s_{NN}}=5.02$ TeV.
The solid graphs are  cross section without nuclear shadowing and
dashed graphs are  with nuclear gluon shadowing parameters.}
\end{figure}
\begin{table}[!ht]
\vspace{-0.4cm}
\caption{
Cross section of photoproduction of $\Upsilon$(1S) in PbPb collisions 
at $\sqrt s= 5.02$ TeV in CMS acceptance
$-2.4<y<2.4$ with different nuclear gluon shadowing parameterizations.}
\label{table:crossPbPb}
\begin{ruledtabular}
\begin{tabular}{|c|c|c|c|c|}
Param.&
\multicolumn{4}{c|}{Gluon shadowing} \\
&\multicolumn{4}{c|}{$\sigma$($\mu$ b)} \\
\cline{2-5}
&w/o shad.&EPS09LO &HKN07LO&\\ 
\hline
CTEQ6L1&166&89&99&\\
\hline
Fit&80&46&49&\\
\hline
\hline
&w/o shad.&EPS09NLO &HKN07NLO&nCTEQ15\\ 
\hline
CTEQ6M&62&37&38 &26\\
\end{tabular}
\end{ruledtabular}
\end{table}
The photoproduction of vector mesons in nucleus-nucleus collisions
is an important tool to study nuclear shadowing of gluon PDFs.
In this section, we  present results  of $\Upsilon$(1s) photoproduction cross section
 for PbPb UPC collisions at ${\sqrt s_{NN}}=5.02$ TeV which is the LHC run 2 scenario.
Fig.~7 shows the rapidity distribution of $\Upsilon$(1s) photoproduction cross section,
with different  nuclear gluon shadowing. 
The solid graphs are  cross section without nuclear shadowing and
dashed graphs are  with nuclear shadowing. 
Figure shows results with LO PDF and LO gluon shadowing from 
EPS09LO~\cite{Eskola:2009} and HKN07LO~\cite{HKN07:2007} 
 and  the results with NLO shadowing parameterizations from EPS09NLO~\cite{Eskola:2009}, HKN07NLO~\cite{HKN07:2007}
and nCTEQ15~\cite{nCTEQ15:2015}.
It is noticed that, nuclear gluon shadowing, 
affects the cross section quite substantially for $\Upsilon$(1S). 
It should be mentioned that, the measurement of $J/\psi$ photoproduction cross section for PbPb 
UPC at 2.76 TeV by ALICE, satisfactorily~\cite{alice01,alice02}
reproduced by moderate gluon shadowing parameterizations from EPS09LO. 
Table~\ref{table:crossPbPb} gives the rapidity integrated $\Upsilon$(1S) 
photoproduction cross section for CMS acceptance ($-2.4<y<2.4$) for PbPb UPC at ${\sqrt s=5.02}$ TeV.
The prediction of cross section from power law fit to HERA+LHCb data
(referred as Fit in Table~\ref{table:crossPbPb}) is also quoted.  
The effect of shadowing is quite prominent in PbPb collisions at 
${\sqrt s_{NN}}= 5.02$ TeV and cross-sections are reduced by $46\%$, 
$40\%$ for CTEQ6L1 gluon PDF with EPS09LO and HKN07LO gluon shadowing respectively.
 The photoproduction cross-section for PbPb collisions with 
CTEQ6M free gluon PDF shows $40\%$, $39\%$ and $58\%$ reduction for EPS09NLO, HKN07NLO
and nCTEQ15 nuclear gluon shadowing parameterizations, respectively.

Our results are also comparable with the earlier
studies of $\Upsilon$(1S) photoproduction.  
The cross section for pPb at ${\sqrt s_{NN}}= 5.02$ TeV, 
 $236$ nb with CTEQ6L1, 
is comparable with earlier estimation with EPS09~\cite{Adeluyi:2013}.
With NLO PDF CTEQ6M, the cross-section  is $146$ nb 
and  from Fit $140$ nb (see Table~\ref{table:crossPbPb}) at ${\sqrt s_{NN}}= 8.16$ TeV
which is comparable to the value of $0.10-0.15$ $\mu$b in Ref.~\cite{Goncalves:2017}
for IIM, bCGC and IP-SAT model predictions.
The cross-sections are marginally different ($6-7\%$) with gluon 
shadowing parameterizations at ${\sqrt s_{NN}}= 5.02$ TeV.
The effect of shadowing is  $7-8\%$ in 
pPb collisions with ${\sqrt s_{NN}}= 8.16$ TeV.
The cross-sections for PbPb collisions 
are also comparable to earlier 
predictions~\cite{Adeluyi:2012aa,Lappi:2013,Guzey:2016,Goncalves:2014,sampaio,Ryskin:2013jmr,Silveira:2015,Goncalves:2017,Silveira:2017}.
In our prediction, we have not considered NLO and other correction factors~\cite{Ryskin:2013jmr}
which may reduce the cross-section and may be the possibile inclusion in future.   
\section{Conclusions}
In the present study, we have considered elastic photoproduction
of $\Upsilon$(nS) in ultraperipheral pPb and PbPb collisions at LHC.
The photoproduction of vector meson is very useful to constrain the
gluon modification, saturation or shadowing, due to the fact that, cross section is quadratically  dependent
on the gluon PDF. The predictions of  pPb cross sections  are consistent 
with HERA, and LHCb data  whereas future measurement will constrain about
the gluon saturation in very low Bjorken $x$. 
The cross section of Upsilon photoproduction in PbPb UPC
are found to be quite sensitive to gluon modifications and  expect to extract information
of nuclear shadowing.
\vspace{0.8cm}

\begin{acknowledgments}
\vspace{-0.6cm}
We would like to express our deepest thanks to Dr. A. K. Mohanty and Dr. David d'Enterria 
for  fruitful discussions. We also acknowledge M. G. Ryskin, P. Jones, A. D. Martin and T. Teubner,
as we used the plot of LO and NLO from Ref.~\cite{Ryskin:2013jmr,CMS:pPb} for comparison
and estimation of rapidity distribution.
\end{acknowledgments}

\nocite{*}

\bibliography{ups_ddutta_arxiv}

\end{document}